\documentclass[10pt,superscriptaddress,twocolumn,amsmath,amssymb,aps,prl,showpacs]{revtex4-1}
\usepackage{mathrsfs}
\usepackage{graphicx}
\usepackage{dcolumn}
\usepackage{bm}
\usepackage[title,titletoc,header]{appendix}

\usepackage{amssymb}
\usepackage{amsmath}
\usepackage{mathrsfs}
\usepackage{amsmath}
\usepackage{subfigure}
\usepackage{float}
\usepackage[title,titletoc,header]{appendix}
\usepackage{graphicx}

\newcommand{\be}{\begin{equation}}
\newcommand{\ee}{\end{equation}}
\newcommand{\bea}{\begin{eqnarray}}
\newcommand{\eea}{\end{eqnarray}}

\begin{document}
\title{Topological Superfluids and BEC-BCS Crossover in Attractive Haldane-Hubbard Model}
\author{Yi-Cai Zhang}
\affiliation{Department of Physics and Center of Theoretical and
Computational Physics, The University of Hong Kong, Hong Kong,
China}

\author{Zhihao Xu}
\affiliation{Institute of Theoretical Physics, Shanxi University, Taiyuan 030006, China}

\author{Shizhong Zhang}
\affiliation{Department of Physics and Center of Theoretical and
Computational Physics, The University of Hong Kong, Hong Kong,
China}

\date{\today}
\begin{abstract}
Motivated by the recent realization of the Haldane model in shaking optical lattice, we investigate the effects of attractive interaction and BEC-BCS crossover in this model at and away from half filling. We show that, contrary to the usual $s$-wave BEC-BCS crossover in the lattice, a topological superfluid with Chern number $C=2$ appears in an extended region of phase space for intermediate strength of the attractive interaction in the interaction-density plane. When inversion symmetry is broken, a new gapless topological state is realized. We also investigate the fluctuations in these superfluid phases and show that the Anderson-Bogoliubov mode is quadratic due to time-reversal symmetry breaking and the existence of an undamped Leggett mode in the strong coupling limit.
\end{abstract}
\pacs{03.75.Kk, 67.85.Lm, 71.10.Fd, 74.20.-z }

\maketitle

{\em Introduction}.
Cold atoms in optical lattices can be used to simulate important models in condensed matter physics due to their high controllability and versatility \cite{Bloch,Lewenstein}. This was demonstrated beautifully by the recent realizations of Haldane \cite{Aoki,Zhai,Jotzu} and Hofstadter models using shaking lattice and Raman laser techniques, respectively. In these experiments, the existence of topological bands is verified using Bloch oscillations with either a Bose condensate~\cite{Stuhl} or free fermions \cite{Aoki}. With tunable interactions in the optical lattices, these advances open new avenues for the controlled study of interaction effects in topological system  and pave the way to the possibility of realizing fractional Chern insulators.

Perhaps by far the best studied interacting topological state is the fractional quantum Hall state, in which the strong Coulomb repulsion between electrons generate emergent fractional quasi-particles that obey abelian or non-abelian statistics \cite{ Laughlin, Zhang,Jain,Wen,Neupert,Tang,Sun}. Naturally, with the discovery of topological insulators \cite{xiaoliang,Kane}, a great deal of efforts has been to investigate its interacting counterparts \cite{Wang,Gurarie,Wu,Hohenadler}. In the case of Haldane model, several recent studies have focused on the interplay between magnetic instabilities and the possible topological ground states in the case of repulsive interactions \cite{Kou,He,Maciejko,Hickey,Zheng}, as well as possible superconducting states with attractive interactions at half filling  \cite{Liang}.

In this Letter, we consider the analogous of BEC-BCS crossover \cite{Zwerger2011} in the attractive Haldane-Hubbard model. In contrast to usual BEC-BCS crossover, we found that away from half filling, there are extended regions of parameter space (interaction-density) for which a topological superfluid is the ground state. With increasing breaking of inversion symmetry, a gapless topological state intervenes before the system enters a trivial superfluid. We also consider the fluctuation effects on the ground states and show that the usual Anderson-Bogoliubov mode becomes quadratic in the absence of time-reversal symmetry and undamped Leggett mode appears in the strong coupling limit. Damping of the collective modes due to coupling to quasi-particles reaches maximum as one approaches phase transition critical point.

{\em The Model}. Consider a two-component Fermi gas with spin $\sigma=\uparrow,\downarrow$, interacting via an onsite attractive interaction $-U$, which can be modeled by the following Haldane-Hubbard model
\be
\mathcal{H}=\sum_{ij\sigma}t_{ij} c_{i\sigma}^\dagger c_{j\sigma}+\sum_{i\sigma}(M\epsilon_i-\mu)n_{i\sigma}-U\sum_in_{i\uparrow}n_{i\downarrow}
\label{eqn:hamiltonian}
\ee
where $\mu$ is the chemical potential. $t_{ij}$ is the hopping amplitude between site $i$ and $j$. For nearest neighbor hopping, $t_{ij}=t$ and will be set equal to one in the following. For the next-nearest-neighbor hopping $t_{ij}=t'\exp(-i\nu \phi)$ with $\nu=1$ for clockwise hopping and $\nu=-1$ for anti-clockwise hopping~\cite{Haldane} (see Fig. 1). Experimentally, $\phi$ is induced by a circular shaking of the optical lattice and except for $\phi=0,\pi$, will break the time-reversal symmetry. $M$ describes the energy offset of the two sub-lattice with $\epsilon_i=1(-1)$ for $A(B)$ sub-lattice and breaks the inversion symmetry of $A$-$B$ sub-lattices. In the following, we use operators $a_i^\dagger$ and $b_i^\dagger$ to denote the fermion creation operator at sub-lattice $A$ and $B$, and $i$ now index the unit cell, which consists of neighboring $A$ and $B$ sites. Now, introducing three Pauli matrix $\boldsymbol{\tau}\equiv (\tau_x, \tau_y, \tau_z)$ that describe the sub-lattice degrees of freedom, the non-interacting Hamiltonian $\mathcal{H}_0$ can be conveniently written in momentum space as  $\mathcal{H}_0({\bf k})=\sum_{\sigma}\psi^\dagger_\sigma({\bf k}) h({\bf k})\psi_\sigma({\bf k})$, where $\psi^\dagger_\sigma({\bf k})=[a_{\sigma}^\dagger({\bf k}),b_{\sigma}^\dagger({\bf k})]$ and $h({\bf k})=\epsilon({\bf k})+{\bf d}({\bf k})\cdot\boldsymbol{\tau}$. $\epsilon({\bf k})=2t'\cos\phi\sum_\delta\cos({\bf k}\cdot{\bf b}_\delta)-\mu$, $d_x({\bf k})=t\sum_\delta\cos({\bf k}\cdot{\bf a}_\delta)$, $d_y({\bf k})=-t\sum_\delta\sin({\bf k}\cdot{\bf a}_\delta)$ and $d_z({\bf k})=M-2t'\sin\phi\sum_\delta\sin({\bf k}\cdot{\bf b}_\delta)$, where ${\bf k}=(k_x,k_y)$ and the vectors ${\bf a}_\delta$ and ${\bf b}_\delta$ are depicted in Fig. 1.

\begin{figure}
\begin{center}
\includegraphics[width=\columnwidth]{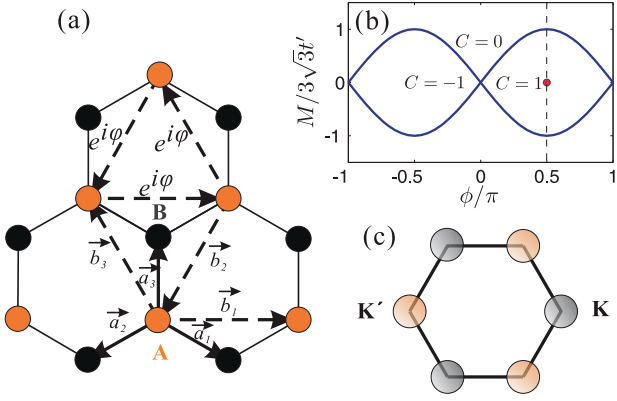}
\end{center}
\caption{(Color online.) The panel (a): the honeycomb lattice in Haldane model. Here two dimensional vectors $\vec{a}_1=[\sqrt{3}/2,-1/2]$, $\vec{a}_2=[-\sqrt{3}/2,-1/2]$ and $\vec{a}_3=[0,1]$. $\vec{b}_1=[\sqrt{3},0]$, $\vec{b}_2=[-\sqrt{3}/2,-3/2]$ and $\vec{b}_3=[-\sqrt{3}/2,3/2]$. The red (black) sites denote sublattice A (B). The phase factor $e^{i\phi}$ associated with next nearest neighbor anti-clock hopping is also labeled. The panel (b) is the Haldane's topological phase diagram in $M-\phi$ plane.  The panel (c) show two Fermi pockets around Dirac points \textbf{$K$} and \textbf{$K'$}.}
\label{schematic}
\end{figure}

For half-filling and with $U=0$, the system is either a trivial insulator for $|M|>3\sqrt{3}t'|\sin\phi|$, or two copies of Chern insulator for $|M|<3\sqrt{3}t'|\sin\phi|$ because of spin degeneracy. Away from half filling, in general, the Fermi surface consists of two (particle or hole) pockets located at $K$ and $K'$ and the system behaves as a metal. It is to be noted that the full Hamiltonian  in Eq.(\ref{eqn:hamiltonian}) respects the $SU(2)$ spin rotation symmetry.

For $\phi=\pi/2$, the Hamiltonian, Eq.(\ref{eqn:hamiltonian}), is also invariant under the particle-hole transformation, $c_{i\sigma}\to \epsilon_i c_{i\sigma}^\dagger$ and $c^\dagger _{i\sigma}\to \epsilon_i c_{i\sigma}$ for $M=0$. For general $M\neq 0$, however, one needs to make further an inversion transformation ${\bf r}\to -{\bf r}$ and an interchange of the $A$ and $B$ sub-lattice, in order that the Hamiltonian remains invariant. Thus, for $\phi=\pi/2$, which will be our focus in the following, it is only necessary to consider the case when $n>2$ (here $n=2$ means two particles per unit cell on average) and the chemical potential satisfies the relation $\mu(-\Delta n)=-\mu(\Delta n)$~\cite{com1}, where $\Delta n\equiv n-2$ is the deviation from half filling. As a result, $\mu$ remains zero at half filling, irrespective of the strength of $U$.

{\em Mean-field phase diagram and Topological Superfluids}. To take into account of the attractive interactions between opposite spins, we make the mean field decoupling in the Cooper channel and introduce $\Delta_i=-U\langle c_{i\uparrow}c_{i\downarrow}\rangle$. In the following, we take the following specific parameters: $t=1$ as energy unit, $\phi=\pi/2$ and $t'=0.15$. We have checked that for $M=0$, $\Delta_i$ is uniform and real, in which case, we shall denote it as $\Delta$. For $M\neq 0$, however, $\Delta_i$ is real but differs for $A$ and $B$ sub-lattice, and we denote them as $\Delta_{\rm A}$ and $\Delta_{\rm B}$, respectively~\cite{com1}. In the later case, the Bogoliubov-de Genne equation takes the form $\mathcal{H}_{\rm BdG}=\sum_{\bf k}\Psi^\dagger({\bf k})H_{\rm BdG}({\bf k})\Psi({\bf k})$, with $\Psi^\dagger({\bf k})=[a_{\uparrow}^\dagger({\bf k}),b_{\uparrow}^\dagger({\bf k}),a_{\downarrow}(-{\bf k}),b_{\downarrow}(-{\bf k})]$ and
\begin{align}
H_{\rm BdG}=\left[\begin{array}{cccc}
h_{11}({\bf k}) &h_{12}({\bf k}) &\Delta_{\rm A} & 0\\
h_{21}({\bf k}) &h_{22}({\bf k}) &0 & \Delta_{\rm B}\\
\Delta_{\rm A} & 0 & -h_{11}^*(-{\bf k}) & -h^*_{12}(-{\bf k})\\
0 & \Delta_{\rm B} & -h_{21}^*(-{\bf k}) & -h^*_{22}(-{\bf k})
  \end{array}\right].
\end{align}
Upon diagonalizing $H_{\rm BdG}$, we obtain the excitation energies given by $E^{p(h)\pm}({\bf k})$, where $\pm$ labels the two branches arising from the underlying $A$-$B$ sub-lattice and $p(h)$ denotes the particle (hole) branch due to particle-hole symmetry of $H_{\rm BdG}$. The particle-hole symmetry of BdG Hamiltonian implies $E^{h\pm}({\bf k})=-E^{p\pm}({\bf -k})$.  The thermodynamic potential (per unit cell) $\Omega$ at zero temperature is then given by $\Omega=\frac{1}{N}\sum_{\bf k}[2\epsilon({\bf k})-|E^{p+}({\bf k})|-|E^{p-}({\bf k})|]+(\Delta_{\rm A}^2+\Delta_{\rm B}^2)/U$~\cite{com1}. The mean field gaps are determined by setting $\partial \Omega/\partial \Delta_{\rm A,B}=0$ and the chemical potential $\mu$ is fixed by requiring the average number of particles to be $n=-\partial\Omega/\partial\mu$.
We solve these three equations self-consistently for $\mu$ and $\Delta_{\rm A,B}$. To further characterize the mean field phase diagram, we also compute the following quantities: (1) the Chern number $C$ \cite{Thouless} corresponding to the two quasi-hole bands with energies $E^{h\pm}({\bf k})$, (2) The existence or not of the edge states in a finite stripe, and (3) The existence or not of the bulk gap $E_{\rm gap}$ in the quasi-particle spectrum.

\begin{figure}
\begin{center}
\includegraphics[width=\columnwidth]{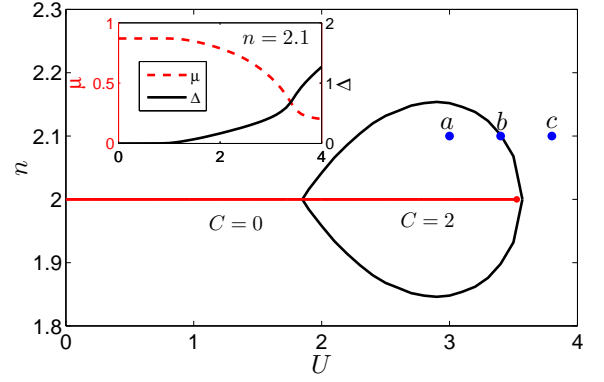}
\end{center}
\caption{(Color online.) (a) Phase diagram for $M=0$ in the $n-U$ plane.  The region within the black line (except the red line) is in topological superfluidity phase with Chern number $C=2$. The exterior region of black line is topologically trivial ($C=0$). The red line denotes the anomalous quantum Hall insulator ($\Delta_{A(B)}\equiv 0$) at half filling ($n=2$). With increasing interaction, {\em e.g.} from point \textbf{a} to \textbf{c}, the system crosses the phase boundary. The inset shows chemical $\mu$ potentials and the pairing gaps ($\Delta_A=\Delta_B=\Delta$) as functions of interaction strength $U$ for filling factor $n=2.1$.  }
\label{pdm0}
\end{figure}

\begin{figure}
\begin{center}
\includegraphics[width=\columnwidth]{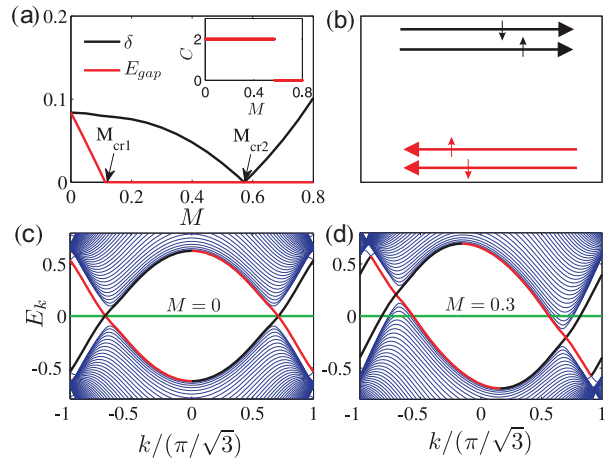}
\end{center}
\caption{(Color online.)  (a): $E_{\rm gap}$ and $\delta$ as a function of $M$ for $U=3.0$ (filling $n=2.1$).  There exists two critical values $M_{\rm cr1}=0.1135$ and $M_{\rm cr2}=0.5718$ . When $M<M_{\rm cr1}$, the bulk gap $E_{\rm gap}$ remains finite, and the system is in the gapped topologically phase. Further increasing $M$ till $M=M_{\rm cr1}$, $E_{\rm gap}$ closes, the system is in the gapless topological phase. For $M>M_{\rm cr2}$, the system becomes  topologically trivial superfluid. For $M_{\rm cr1}<M<M_{\rm cr2}$, the system is in the gapless topological superfluid state. The Chern number is shown in inset. (b) Edge states in the boundary of strip geometry. (c)  quasi-particle spectrum (blue) and edge states (red (black) for left (right) going  ) in gapped topological phase for $M=0$. (d) quasi-particle spectrum (blue) and edge states (red (black) for left (right) going ) in gappless topological phase for $M=0.3$.}
\label{pdmn0}
\end{figure}

(i) $M=0$. In this case, inversion symmetry is respected and $\Delta_i\equiv\Delta$ is uniform. The phase diagram in the $n$-$U$ plane is shown in Fig. \ref{pdm0}. At half filling, the system is a Chern insulator with $C=2$ due to spin degeneracy for small attractive interaction (red line in Fig. \ref{pdm0}). At a critical value $U\approx3.526$, the system enters a topological superfluid state with $C=2$ which, however, occupies only a very small parameter regime before it enters the trivial superfluid, with $C=0$ (crossing phase boundary at $U\approx3.57$). Above half filling $n>2$, the situation changes dramatically. The non-interacting system is metallic with two Fermi surfaces located around $K$ and $K'$ points, which become superfluid due to attractive interactions.  When $U$ is either small or large, the superfluid state is trivial and characterized by $C=0$. However, for intermediate attractive interactions, the system enters a topological superfluid state with $C=2$ in a significantly enlarged portion of the phase space as compared to that at half filling. Inside the topological phase, a bulk gap is always present. In a finite stripe, a pair of edge states with opposite spin exist on both sides of the stripe and carries a net charge current but no spin current, as shown in inset of Fig.~\ref{pdmn0} (b).

The main feature of the mean field phase diagram can be understood from the following observations. It is well known that the topological transition is accompanied by the change of the topology of the energy band and at the transition point, energy gap will close. In the case when $M=0$, the band closing occurs when $3\sqrt{3}t'=\sqrt{\mu^2+\Delta^2}$ at $K$ and $K'$ points. As one increases the interaction $U$, $\Delta$ increases but $\mu$ decreases (see inset of Fig.\ref{pdm0}). As a result, in the intermediary value of $U$, the condition $3\sqrt{3}t'>\sqrt{\mu^2+\Delta^2}$ is satisfied and a $C=2$ topological state emerges.

{(ii)}  $M\neq 0$. In this case, the inversion symmetry is broken and $\Delta_{\rm A}\neq \Delta_{\rm B}$. In addition to the trivial superfluid state for which $C=0$, there appear two different topological superfluid states. When $|M|$ is small, a fully gapped topological state appears with $C=2$ [Fig.\ref{pdmn0} (a)], much the same as the case for $M=0$ with a pair of edge states [Fig.\ref{pdmn0} (c)]. As one increases $M$, a gapless topological superfluid state appears when $M=M_{\rm cr1}$, for which the bulk gap $E_{\rm gap}$ becomes zero, while the gaps at $K$ and $K'$ remain finite. In this case, the edge state remains. Further increasing $M$, the system enters into a trivial superfluid state when $M=M_{\rm cr2}$, quasi-particle gap $\delta$ at $K$ or $K'$ close and edge states disappear.

Explicitly, for $M>0$, the closing of the bulk gap $E_{\rm gap}={\rm min}_{k}|E^{p-}(k)|$ that determines $M_{\rm cr1}$ is given by
\be
3\sqrt{3}t'-\sqrt{(\mu+M_{\rm cr1})^2+\Delta_{\rm B}^2}=0.
\ee
For $M<0$, one only needs to replace $\Delta_{\rm B}$ with $\Delta_{\rm A}$. Further increasing $M$, the gaps at $K$ or $K'$ will close. The critical value of $M_{\rm cr2}$ is given by setting $\delta=0$ with
\be
\delta\equiv 3\sqrt{3}t'-\frac{\sqrt{(\mu+M_{\rm cr2})^2+\Delta_{\rm B}^2}+\sqrt{(\mu-M_{\rm cr2})^2+\Delta_{\rm A}^2}}{2}.
\label{eq:criterionM}
\ee
For $M<0$, we only need to interchange $\Delta_{\rm A}$ and $\Delta_{\rm B}$ in Eq.(\ref{eq:criterionM}). The calculated Chern number $C$ as a function of $M$ are shown in the inset of Fig.\ref{pdmn0} (a), where it remains at $2$ for $M<M_{\rm cr2}$.

In the absence of inversion symmetry, it is possible for system to develop the Fulde-Ferrell-Larkin-Ovchinnikov (FFLO) pairing state~\cite{Fulde,Larkin}, in which the order parameter $\Delta_i$ modulates periodically in real space \cite{Xu,Cao,Hu}. We have carried out inhomogeneous Bogoliubov-de Genne calculation in real space, and find no indication of its existence for our chosen parameters~\cite{com1}. We note that the $SU(2)$ spin rotational symmetry is respected, while the time reversal symmetry is broken for our model. Accordingly, the topological superfluidity phases found here belong to class {\bf C} in BdG class, which are characterized by an even Chern number~\cite{Ludwig1,Ludwig2}.


\begin{figure}
\begin{center}
\includegraphics[width=1\columnwidth]{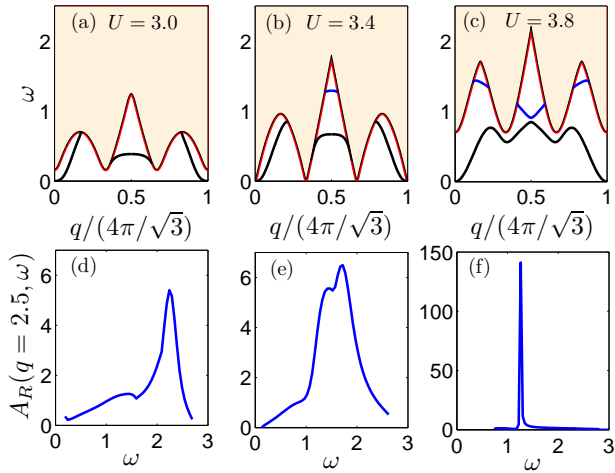}
\end{center}
\caption{(Color online). The collective excitation spectrum for various values of interaction strength ($M=0$): (a) $U=3.0$, (b) $U=3.4$ and (c) $U=3.8$, corresponding to the three blue points \textbf{a}, \textbf{b} and \textbf{c} in  Fig.~\ref{pdm0}, respectively. The momentum ${\bf q}$ is chosen along $\hat{x}$-direction, ${\bf q}=q\hat{x}$. The shaded region denotes the quasi-particle continuum with its lowest boundary (red) given by ${\rm min_k}[E^{-}_{k}+E^{-}_{k+q}]$. There are two types of collective excitations. The black line corresponds gapless Goldstone mode, while the blue line is the Leggett mode.  (d-f) The spectral function of the Leggett mode for various values of interaction strength corresponding to that in (a-c). Note that the maximal damping occurs at the transition point (b,e).}
\label{collective}
\end{figure}

{\em Goldstone mode, Leggett mode and its damping}. We now turn to the discussions of collective modes of the attractive Haldane-Hubbard model. For simplicity, we shall only consider the situation when $M=0$ and calculate the collective modes by considering Gaussian fluctuations around the mean field ground state for $n=2.1$, with two Fermi surfaces around $K$ and $K'$ points (see Fig.\ref{schematic}). To go beyond mean field theory and calculate the mode frequencies, we expand the order parameters: $\Delta_{\rm A,B}(q)=\Delta_{\rm A,B}+\eta_{\rm A,B}(q)$ and integrate out the fermions to arrive at the effective action for the order parameter fluctuation $S_{\rm eff}=\frac{1}{2}\sum_{q}\hat{\eta}^*(q)\hat{M}(q)\hat{\eta}(q)$, with $q=({ \vec{q}},i\omega_n)$ and $\omega_n$ is the bosonic Matsubara frequencies. $\hat{\eta}^{*}(q)=(\eta_{\rm A}^*(q),\eta_{\rm B}^*(q),\eta_{\rm A}(-q),\eta_{\rm B}(-q))$ defines the fluctuation vector in ${\rm A,B}$ sublattice and $\hat{M}(q)$ is the fluctuation matrix whose explicit form is given in the supplementary material~\cite{com1}. The excitation energy is determined by the zeros of its determinant ${\rm Det}|M(\vec{q},i\omega_n\rightarrow \omega+i0^+)|=0$ \cite{Melo}.

Because of the existence of two Fermi surfaces, in addition to the usual Goldstone (or Anderson-Bogoliubov) mode, which corresponds to the oscillation of total density, there appears additional Leggett mode \cite{Paramekanti}, which corresponds to the oscillation of the relative densities of the two Fermi pockets located at $K$ and $K'$ points. In Fig. \ref{collective} (a-c), we show the excitation spectrums for three different values of $U$, corresponding to a $C=2$ topological superfluid (point \textbf{a} in Fig.2), at the phase boundary (point \textbf{b} in Fig.2) and inside the trivial superfluid (point \textbf{c} in Fig.2). The shaded region corresponds to the quasi-particle continuum, which touches zero at the transition point, as expected for a quantum phase transition from topological to trivial superfluid state. In addition, there are a few further notable features of the collective modes.

Firstly, we note that the long wave length Goldstone excitation has a quadratic dispersion $\omega({\bf q})=c|{\bf q}|^2$, unlike the usual linear dependence, typical of superfluid \cite{Melo,Diener}. This is because for the Haldane-Hubbard model, Eq.(\ref{eqn:hamiltonian}), time-reversal symmetry is broken, so the low energy effective theory describing the total phase fluctuation contains the term $\partial_t\varphi$, with $\varphi$ the corresponding fluctuation field. On the other hand, for $M=0$, inversion symmetry is respected and so the spatial derivative must be of the form $\boldsymbol{\nabla}^2\varphi$. Consequently, one finds $\omega({\bf q})=c|{\bf q}|^2$ at long wave length. Furthermore, while the Goldstone mode is always present when $|{\bf q}|\to 0$, it can merge into the quasi-particle continuum and ceases to be a well-defined elementary excitation at larger $|{\bf q}|$ (see Fig. 4 (a and b)).

Secondly, we observe that at strong coupling ($U=3.8$), well-defined Leggett mode appears in the first Brillouin zone, while at weak coupling, it merges into the continuum and damps away. To investigate the damping of Leggett mode when it merges into the quasi-particle continuum, we calculate the spectral function of the relative phases fluctuations, $A_{\rm R}(q)=1/\pi{\rm Im}\langle\theta^*(q)\theta(q)\rangle$, where $\theta(q)\equiv \theta_{\rm A}(q)-\theta_{\rm B}(q)$ and $\theta_{\rm A,B}(q)={\rm Im}\eta_{\rm A,B}$~\cite{com1}. In Fig.(\ref{collective} d-f), we show $A_{\rm R}(|{\bf q}|=2.5, \omega)$ and found that the damping is maximum at the quantum critical point. The Leggett mode can be excited and detected by a differential modulation of the $A$-$B$ sub-lattice~\cite{Paramekanti}.

{\em Discussions}. With the recent realization of Haldane model in cold atoms, it is possible now to study the topological superfluid states and their quantum phase transitions discussed here. Unlike the non-interacting Haldane model in which the quantum phase transitions can be monitored by using Block oscillation and the band mapping technique, here the phase transition is signalled by the closing of quasi-particle gap which will have dramatic effects on the dynamic structure factor that can be measured experimentally with Bragg spectroscopy. The Leggett modes in multi-band superconductor MgB$_2$ have been observed experimentally by tunneling spec troscopy techniques~\cite{Ponomarev2004}, Raman spectroscopy~\cite{Blumberg2007}, and angle-resolved photoemission spectroscopy~\cite{Mou2015}. Here we expect that for neutral atoms, the Leggett modes and their damping can be measured by the Bragg spectroscopy or modulation spectroscopy.

\emph{Acknowledgements.} We would like to thank Bohm-Jung Yang for useful discussions. This research is supported by Hong Kong Research Grants Council (General Research Fund, HKU 17306414 and Collaborative Research Fund, HKUST3/CRF/13G) and the Croucher Foundation under the Croucher Innovation Award.

\end{document}